\documentclass[%
 reprint,
nofootinbib,
 amsmath,amssymb,
 aps,
prd,
]{revtex4-2}

\usepackage{graphicx}
\usepackage{dcolumn}
\usepackage{bm}
\usepackage{hyperref}
\usepackage[normalem]{ulem}

\usepackage{color}
\usepackage{multirow}

\begin{document}

\preprint{APS/123-QED}

\title{Nonradial instabilities in anisotropic neutron stars}

\author{Shu Yan Lau}
\email{sl8ny@virginia.edu}
\affiliation{Department of Physics, University of Virginia, Charlottesville, Virginia 22904, USA}
\author{Siddarth Ajith}
\affiliation{Department of Physics, University of Virginia, Charlottesville, Virginia 22904, USA}
\author{Victor Guedes}
\affiliation{Department of Physics, University of Virginia, Charlottesville, Virginia 22904, USA}
\author{Kent Yagi}
\email{ky5t@virginia.edu}
\affiliation{Department of Physics, University of Virginia, Charlottesville, Virginia 22904, USA}




\date{\today}

\begin{abstract}
Non-radial oscillation modes of a neutron star possess valuable information about its internal structure and nuclear physics. Starting from the quadrupolar order, such modes under general relativity are known as quasi-normal modes since they dissipate energy through gravitational radiation and their frequencies are complex. The stability of these modes is governed by the sign of the imaginary part of the frequency, which determines whether the mode would decay or grow over time.
In this paper, we develop a fully consistent framework in general relativity to study quasi-normal modes of neutron stars with anisotropic pressure, whose motivation includes strong internal magnetic fields and non-vanishing shear or viscosity. We employ parametrized models for the anisotropy, with a specific form under perturbation that is consistent with previous studies under the Cowling approximation, and solve the perturbed Einstein field equations numerically. The assumed form of the perturbed anisotropy can be easily extended to more general cases. We find that, unlike the case for isotropic neutron stars, the imaginary parts of some of the pressure ($p$-)modes flip signs as the degree of anisotropy deviates from zero, depicting a transition from stable modes to unstable modes. This finding indicates that some anisotropic neutron star models are unstable, potentially restricting the form of sustained anisotropy. 

\end{abstract}

\maketitle


\section{\label{sec:intro} Introduction}

Pulsation modes provide the basis to study the dynamical deformations of a neutron star (NS). The frequency of these modes characterizes the stellar internal composition and density profile. In addition, the stability of the modes also helps predict whether certain configurations of a star can exist in nature. With the advancement in multimessenger astronomy over the years, the detection of some of the pulsation modes is expected in the future \cite{Tsang_2012, Chirenti_2017, Yu_2017, Pratten_2020, Pan_2020, Ng_2021, Lau_2021, Passamonti_2021}.

Within Newtonian theory, the adiabatic pulsation modes are normal modes described by real frequencies in the absence of dissipation, radiation reaction, and various types of instabilities (see, e.g., \cite{Cox_1958, Chandrasekhar_1967, Wood_1976, Friedman_1978, Cox_1980, Gautschy_1990}).
For even parity (polar) non-radial modes, the well-known ones include the fundamental ($f$-)modes, the pressure ($p$-)modes, and the gravity ($g$-)modes. In the original classification by Cowling \cite{Cowling_1941}, the naming of these modes is based on the local propagation behavior of the waves that form the modes. For example, the $p$-modes, which are of interest in this study, can be conceived as the standing waves formed by propagating acoustic waves within the bulk via a local analysis \cite{Gabriel_1979, Unno_1979, Cox_1980}. 

Under general relativity (GR), non-radial oscillation modes (except the dipole modes) can dissipate energy through the emission of gravitational radiation. These modes are described by oscillations with complex frequencies, denoted by $\omega$ hereafter, and are known as the quasinormal modes (QNMs). The complex values are the characteristics of an open system. Despite many of those being referred to as the `fluid modes' \cite{Kokkotas_1999}, which are analogous to the Newtonian stellar pulsation modes, they are fundamentally different due to the boundary conditions in the exterior vacuum region of the star. 

The stability of the QNMs is determined by the sign of the imaginary part \cite{Thorne_1967_1}. For our sign convention of the time dependence of the QNMs expressed as $e^{i\omega t}$, the mode will grow exponentially over time if the imaginary part is negative, thus being unstable.

Typical NSs are described as \emph{isotropic} fluids, in which the local stress is invariant under rotational transformations. Stability analysis on the QNMs of non-rotating isotropic fluid stars has shown that the unstable modes must have pure imaginary frequencies, i.e., the mode must be non-propagating \cite{Detweiler_1973, Ipser_1975, Lebovitz_1981}. 
This stability criterion can be extended to certain regimes, including the two-fluid NS models \cite{Wong_2009}, where the argument given in \cite{Detweiler_1973} still applies.
A similar condition has also been shown to hold in the scalar-tensor theories for the radial scalar-modes \cite{Harada_1997, Mendes_2018}, an example of QNMs in non-GR theories. 

Meanwhile, local anisotropy can exist within a NS in various scenarios involving elasticity \cite{Heintzmann_1975, Karlovini_2003, Nelmes_2012}, superfluidity \cite{Letelier_1980, Herrera_1997}, pion condensation \cite{Sawyer_1973}, strong magnetic field \cite{Cardall_2001, Huang_2010, Ferrer_2010, Yazadjiev_2012, Sinha_2013}, and viscosity \cite{Barreto_1992}. Certain exotic relativistic stars are also predicted to have anisotropic stress, like boson stars \cite{Kaup_1968, Schunck_2003, Alcubierre_2022}, strange stars \cite{Deb_2017, Errehymy_2019}, dark matter stars \cite{Moraes_2021}, and gravastars \cite{Cattoen_2005, Chirenti_2007}. 
The solution to a static spherically symmetric anisotropic NS was first studied in \cite{Bowers_1974}. After that, much work has been done on calculations of the NS structure with different anisotropic equations of state (EOSs) as well as assessing the radial stability \cite{Bayin_1982, Hillebrandt_1976, Dev_2003, Karlovini_2004, Horvat_2010, Horvat_2011, Arbail_2016, Isayev_2017, Pretel_2020}.  One interesting finding is that depending on the anisotropy analysis, the maximum-mass configuration of the anisotropic NSs may or may not correspond to a change in radial stability \cite{Arbail_2016, Isayev_2017, Pretel_2020}, in contrast to the isotropic NSs \cite{Bardeen_1966}. 

The non-radial deformations of anisotropic NSs are also studied in previous literature. In the static limit, such perturbations within the slow rotation and small tidal deformation approximation allow one to calculate the moment of inertia, spin-induced quadrupole moment, and tidal deformability \cite{Silva_2015, Yagi_2015} that are known to enjoy some universal relations~\cite{Yagi:2013bca,Yagi:2013awa,Yagi:2016bkt}. 
Anisotropic NSs were used to also study how universal relations for multipole moments of NSs approach the black hole limit~\cite{Yagi:2015upa,Yagi_2016}. For non-radial time-dependent perturbations, Refs. \cite{Doneva_2012, Curi:2022nnt, Mohanty_2023, Jyothilakshmi_2024} employed the relativistic Cowling approximation in calculating the pulsation modes of anisotropic NSs by ignoring the metric perturbations. Such an approximation can accurately predict the frequencies of non-radial oscillation modes for isotropic stars with an error of $< 10\%$ for $p$-modes, which decreases with the order of the mode \cite{Lindblom_1990}, while its validity is currently unknown for anisotropic stars.  
Reference \cite{Hillebrandt_1976} has also investigated the non-radial instability of anisotropic stars, demonstrating that the non-radial pulsation modes are stable except for infinitely large anisotropy. They, however, considered only the Newtonian limit of incompressible stars for a specific anisotropy ansatz that turns out to be unphysical as it causes a divergence at the stellar center.

In this work, we develop a consistent framework to treat stellar perturbations for anisotropic fluids in full GR. Although there are many works studying stellar perturbations for isotropic stars in full GR since the early 1970s~\cite{Detweiler_1973, Ipser_1975, Lebovitz_1981}, there was no such formulation available for anisotropic stars for more than 50 years. The new formulation developed here allows us to check the validity of the Cowling approximation and study the stability of anisotropic stars for the first time.
Our numerical result shows that the anisotropy causes certain $p$-modes to acquire a negative imaginary part in the eigenfrequencies while having a non-zero real part. This can be interpreted as an oscillatory mode with exponentially growing amplitude, which is in stark contrast with the case for isotropic NSs. We further demonstrate how the mathematical proof of the non-existence of oscillatory unstable modes for isotropic NSs does not apply to the anisotropic case by considering a similar integral formula as Eq.~(16) of \cite{Detweiler_1973}. We also found that the Cowling approximation is valid for anisotropic NSs with an error of $< 30\%$ for $f$-modes and $< 10\%$ for $p$-modes  from the fully consistent analysis.

The rest of this paper is organized as follows: In Sec.~\ref{sec:aniso_NS}, we describe the two anisotropy models we employ in this study. Next, we give the formalism for computing the non-radial pulsations of a spherically symmetric non-rotating anisotropic NS in Sec.~\ref{sec:formulation}. We then show the results of the mode instability in Sec.~\ref{sec:unstable_modes}. Lastly, we discuss the interpretations of the results and possible extensions of the work in Sec.~\ref{sec:conclusion}.

Unless otherwise specified, we employ the geometrized unit system with $G = c = 1$ in this paper.

\section{\label{sec:aniso_NS} Anisotropic neutron stars}
The fluid stress within an anisotropic NS has directional dependence. The stress-energy tensor describing the anisotropic fluid with spherical symmetry takes the form \cite{Bowers_1974}
\begin{align}
    T_{\alpha \beta} = \rho u_\alpha u_\beta + p_r h_{\alpha \beta} - \sigma \Omega_{\alpha \beta},
\end{align}
where $\rho$ is the energy density, $\sigma = p_r - p_t$ is the degree of anisotropy, $p_r$ and $p_t$ represent the pressures in the radial and tangential direction respectively, $h_{\alpha \beta} = g_{\alpha\beta}+u_\alpha u_\beta$ is the transverse metric on a 3D space, $\Omega_{\alpha \beta} = h_{\alpha\beta}-k_\alpha k_\beta$ is another transverse metric on a 2D sphere, $u^\alpha$ is the four-velocity vector of a fluid, and $k^\alpha$ is the unit normal vector in the radial direction orthogonal to $u^\alpha$.

In this paper, we consider two anisotropy ansatzes described by quasi-local parameters. These anisotropy models contain not only the local thermodynamic parameters, like $\rho$, or $p_r$, but also quasi-local parameters that depend on the geometry at the specified point of the spacetime. The first one is modified from the anisotropy model given in Horvat \emph{et al.} \cite{Horvat_2010}:
\begin{align}
\label{eq:H-model}
    \sigma = \beta p_r \mu^2,
\end{align}
where $\mu = 2 m/r$, $m$ is the gravitational mass enclosed within a radius $r$, and $\beta$ is a dimensionless parameter governing the degree of anisotropy. We denote this ansatz as the `H-model'. Note that compared to \cite{Horvat_2010}, $\sigma$ here contains an extra factor of $\mu$. Such an additional factor is introduced to satisfy the regularity conditions of the perturbations at $r = 0$, which requires $\partial \sigma/ \partial \mu = \mathcal{O}(r^2)$ as shown in Eq.~\eqref{eq:dsigma_regular}. The above model still satisfies the conditions used to construct the original model, namely (i) the anisotropy term to vanish from the hydrostatic equilibrium equation in the non-relativistic limit ($p_r \ll \rho$), and (ii) the anisotropy vanishes at the stellar center. Due to the additional factor of $\mu$ in the new model which is of the order $\sim0.1$, we use the range $|\beta|\le 10$ instead of $|\beta|\le 2$ in the original model. Note that in the relativistic Cowling approximation used in \cite{Doneva_2012}, the original anisotropy model satisfies the regularity conditions automatically since there are no metric perturbations.

The second model is proposed by Bowers and Liang \cite{Bowers_1974}:
\begin{align}
    \sigma = \beta \frac{(\rho+p_r)(\rho+3p_r)}{1-\mu} r^2,
\end{align}
where $|\beta|\le 0.5$ as in \cite{Pretel_2020}.
This anisotropy model permits an analytic solution to the stellar structure for incompressible stars. We denote it as the `BL-model'. Unlike the H-model, we do not need to modify the BL-model as it satisfies the regularity conditions at the center.

The relation between $\rho$ and $p_r$ is provided by an EOS. We choose the polytropic EOS named `EOS II' in \cite{Doneva_2012} (denoted by `Poly' hereafter) and a more realistic EOS, SLy4 \cite{Douchin_2001}, to demonstrate our major numerical results. 
With these anisotropy ansatzes and EOSs, we numerically integrate the structural equations (see, e.g., \cite{Bowers_1974}) to construct the anisotropic NS models at equilibrium.
The stress-energy tensor of all models considered satisfies the dominant, weak, and strong energy conditions. Moreover, the speeds of sound in the radial and tangential directions within the stars (see, e.g., \cite{Becerra_2024, Mohanty_2023}) both satisfy causality. We also check that both $p_r$ and $p_t$ are positive. Note that for some models with large positive $\beta$ outside the range specified above, $p_t$ can become negative near the stellar surface, which is defined at the radius where $p_r = 0$.

\section{\label{sec:formulation} Formulation}
Here we provide the full GR formalism for computing the QNMs for the first time. We construct the background non-rotating solution of an anisotropic star and introduce non-radial time-dependent small perturbations to the metric and fluid variables. We numerically solve the equations to obtain the QNM frequencies and eigenfunctions. For the perturbations, we assume they follow a specific form (given in Eq.~\eqref{eq:dsigma} described below), which reduces to the Cowling approximation \cite{Doneva_2012} when we take the metric perturbations to zero. 
Note that this is not derived from the fundamental properties of anisotropic fluids.
One important assumption of this model is that the perturbed tangential pressure is always the same for both directions along the 2-sphere. We suggest a further generalization in Appendix~\ref{app:continuity_eq} but leave it for future work.

We first define the static spherically symmetric background metric as
\begin{align}
    ds^2 = -e^{\nu}dt^2 + e^{\lambda}dr^2 + r^2 d\Omega^2,
\end{align}
where $\nu$ and $\lambda$ are functions of $r$ while $d\Omega$ is the unit 2-sphere line element. The Einstein field equations of the anisotropic fluid background give (e.g., \cite{Bowers_1974})
\begin{align}
    \nu^\prime =& 2\frac{m+4\pi r^3 p_r}{r^2} e^\lambda, \\
    p_r^\prime =& -(\rho+p_r)\frac{\nu^\prime}{2} - \frac{2\sigma}{r},\label{eq:TOV_p}\\
    \lambda^\prime =& 2\frac{-m+4\pi r^3 \rho}{r^2} e^\lambda, \\
    e^\lambda =& \left(1-\frac{2m}{r}\right)^{-1},
\end{align}
where the superscript prime denotes derivative with respect to $r$.

In the pulsating star, the metric and the matter dynamics are governed by the perturbed Einstein field equations and the stress-energy conservation:
\begin{align}
    \delta G_{\alpha\beta} = 8\pi \delta T_{\alpha \beta}, \label{eq:pEFE}\\
    \delta (\nabla_\alpha T^{\alpha\beta}) = 0, \label{eq:pSEC}
\end{align}
where $\nabla_\alpha$ denotes the covariant derivative, $G_{\alpha \beta}$ is the Einstein tensor and $T_{\alpha \beta}$ is the stress-energy tensor. The prefix $\delta$ represents Eulerian perturbations and is related to the Lagrangian perturbations $\Delta$ through the Lie derivative \cite{Friedman_1975}, $\Delta = \delta + \mathcal{L}_\xi$, given the Lagrangian displacement vector $\xi^\alpha$ of the fluid.

We employ the same convention of the perturbed metric as in \cite{Lindblom_1983, Detweiler_1985}, defined in the Regge-Wheeler gauge \cite{Regge_1957}:
\begin{align}
    \delta g_{\alpha\beta} dx^\alpha dx^\beta =& -\sum_{\ell, m} \big[e^\nu r^\ell H_0 dt^2 +2i\omega r^{\ell+1} H_1 dtdr \nonumber\\
    &+ e^{\lambda} r^\ell H_2 dr^2+  r^{\ell+2} K d\Omega^2\big] e^{i\omega t}Y_{\ell m}. \label{eq:pmetric}
\end{align}
Here $\left(H_0, H_1, H_2, K\right)$ are functions of $r$, and $Y_{\ell m}$ are the spherical harmonics. The radial parts of the expansion depend on the index $\ell$, which is suppressed for conciseness in the following expressions. This applies to all the radial parts of the expansions in spherical harmonics. Since we focus on non-rotating stars with spherical backgrounds, perturbations with different $\ell$ do not couple with each other.

The stress-energy tensor takes the same form as \cite{Doneva_2012}. One major assumption about the perturbed density and radial pressure is the adiabatic relation
\begin{align}
    \Delta p_r = \frac{\gamma p_r}{\rho + p_r} \Delta \rho, \label{eq:pGamma}
\end{align}
where $\gamma$ is the adiabatic index, and is assumed to be 
\begin{align}
    \gamma= \frac{\rho + p_r}{p_r}\left(\frac{\partial p_r}{\partial \rho}\right)_{s} = \frac{\rho + p_r}{p_r}\left(\frac{\partial p_r}{\partial \rho}\right)_\text{eq}.
\end{align}
The subscript `$s$' represents an adiabatic derivative while the subscript `eq' denotes the derivative of the background $p_r$ against $\rho$. This essentially means the EOS remains a single variable function in the dynamical timescale. The Lagrangian displacement vector is defined as
\begin{align}
    \xi^\alpha \partial_\alpha=& \sum_{\ell, m} r^{\ell-1} e^{i\omega t}\Big[ e^{-\lambda/2}WY_{\ell m} \partial_ r \nonumber\\
    &- \frac{V}{r}\left(\frac{\partial Y_{\ell m}}{\partial \theta}  \partial_\theta + \csc^2 \theta \frac{\partial Y_{\ell m}}{\partial \phi} \partial_\phi \right)\Big],\label{eq:displacement}
\end{align}
where $V$ and $W$ are functions of $r$.

Using Eqs.~\eqref{eq:pEFE}--\eqref{eq:displacement}, we arrive at the set of equations governing the stellar pulsations:
\allowdisplaybreaks
\begin{widetext}
\begin{align}
    H_1^\prime =& \left[4\pi(\rho-p_r) e^\lambda r-\frac{2 m e^\lambda}{r^2} - \frac{\ell+1}{r}\right] H_1 + \frac{e^\lambda}{r} K + \frac{e^\lambda}{r}H_0 - \frac{16\pi(\rho+p_r)}{r}e^\lambda\left(1-\bar\sigma\right) V, \label{eq:LD_H1}\\
    K^\prime =& \frac{\ell(\ell+1)}{2r} H_1 +\left(\frac{\nu^\prime}{2} - \frac{\ell+1}{r}\right) K -8\pi(\rho+p_r)\frac{e^{\lambda/2}}{r} W +\frac{1}{r} H_0, \label{eq:LD_K}\\
    W^\prime =& re^{\lambda/2}(1-\bar\sigma) K +\left(- \frac{\ell+1}{r}+\frac{2 \bar\sigma}{r} \right) W + \frac{r e^{(\lambda-\nu)/2}}{\gamma p_r} X + \frac{r e^{\lambda/2}}{2} H_0 -\frac{\ell(\ell+1)}{r} e^{\lambda/2} (1-\bar\sigma) V, \label{eq:LD_W}\\
    X^\prime =& \frac{\rho+p_r}{2} e^{\nu/2}\left[r\omega^2e^{-\nu} + \frac{\ell(\ell+1)}{2r}(1-2\bar\sigma)\right] H_1 + \frac{\rho+p_r}{2} e^{\nu/2}\left[\left(\frac{3}{2}-2\bar\sigma\right)\nu^\prime-\left(1-6\bar\sigma\right)\frac{1}{r} - \frac{4\bar\sigma^2}{r}\right] K \nonumber\\
    &-\frac{\rho+p_r}{r} e^{(\lambda+\nu)/2}\left[4\pi(\rho+p_r) + \omega^2 e^{-\nu} - F \right] W -\frac{1}{r}\left(\ell-2\frac{\rho+p_r}{\gamma p_r} \bar\sigma\right) X \nonumber\\
    &+ \frac{\rho+p_r}{2} e^{\nu/2}\left(\frac{1}{r}-\frac{\nu^\prime}{2}\right) H_0 + \frac{\ell(\ell+1) e^{\nu/2} p_r^\prime }{r^2}(1-\bar\sigma) V + \frac{2e^{\nu/2}}{r}S. \label{eq:LD_X}
\end{align}
\end{widetext}
Here, the overbarred anisotropy parameter is defined by 
\begin{equation}
    \bar\sigma = \frac{\sigma}{\rho+p_r},
\end{equation}
while $X$ is the radial part of the expansion
\begin{equation}
    \sum_{\ell, m} r^\ell X Y_{\ell m} e^{i\omega t} = -e^{\nu/2} \Delta p_r,
\end{equation}
and $F$ in Eq.~\eqref{eq:LD_X} is given by
\begin{widetext}
    \begin{align}
        F = e^{-\lambda/2}\left\{\frac{r^2}{2}\left(\frac{e^{-\lambda/2}\nu^\prime}{r^2}\right)^\prime - e^{-\lambda/2} \left[\left(\frac{6}{r^2}-\frac{2\nu^\prime}{r}\right)\bar\sigma -\frac{2}{r^2}\left(\frac{r\sigma^\prime}{\rho+p_r}\right)-\frac{4}{r^2}\bar\sigma^2\right]\right\}.
    \end{align}
\end{widetext}
The Eulerian perturbation of the anisotropy parameter, $S$, is defined by
\begin{equation}
     \sum_{\ell, m} r^\ell S Y_{\ell m} e^{i\omega t} =  \delta \sigma,
\end{equation}
and is related to other perturbation variables by 
\begin{widetext}
\begin{align}
    S =& -\left[\left(\frac{\partial \sigma}{\partial p_r}\right)_\text{eq} + (\rho+p_r)\left(\frac{A}{p_r^\prime} + \frac{1}{\gamma p_r}\right)\left(\frac{\partial \sigma}{\partial \rho}\right)_\text{eq}\right]\frac{e^{-\lambda/2}p_r^\prime}{r}W \nonumber \\
&- \left[\left(\frac{\partial \sigma}{\partial p_r}\right)_\text{eq} + \frac{\rho+p_r}{\gamma p_r}\left(\frac{\partial \sigma}{\partial \rho}\right)_\text{eq}\right]e^{-\nu/2}X - \left(\frac{\partial \sigma}{\partial \mu}\right)_\text{eq}e^{-\lambda}H_0. \label{eq:dsigma} 
%
\end{align}
\end{widetext}
Here, $A$ is the relativistic Schwarzschild discriminant, which we take to be zero due to our assumption on $\gamma$ (i.e., assuming the EOS is a single variable function of $p_r$), namely
\begin{align}
     A \equiv \frac{\rho^\prime}{\rho+p_r}-\frac{p_r^\prime}{\gamma p_r}  = 0.
\end{align}
We further assume the partial derivatives on $\sigma$ follow the equilibrium background \cite{Doneva_2012}, similar to the case of $\gamma$. In general, these derivatives should depend on the thermodynamic relations in the perturbation timescale \cite{Thorne_1967_2}. 
The Einstein field equations also give $H_2=H_0$ as in the isotropic case. 
The variables ($H_0$,$V$) are related to the dependent variables through the algebraic relations
\begin{widetext}
\begin{align}
    \left[3m + \frac{(\ell-1)(\ell+2)}{2}r +4\pi r^3 p_r\right]H_0 =& 8\pi r^3 e^{-\nu/2} X - \left[ \frac{\ell(\ell+1)}{2}(m+4\pi r^3 p_r) - \omega^2 r^3 e^{-\lambda-\nu}\right] H_1 \nonumber\\
     &+ \left[\frac{(\ell-1)(\ell+2)}{2}r - \omega^2 r^3 e^{-\nu} - \frac{e^\lambda}{r}(m+4\pi r^3 p_r) (3m-r+4\pi r^3 p_r)\right]K \nonumber \\
& -16\pi r e^{-\lambda/2} (\rho+p_r) \bar\sigma W, \label{eq:LD_H0} \\
     \quad\quad X = & \omega^2(\rho+p_r) e^{-\nu/2} (1-\bar\sigma) V + \frac{\rho+p_r}{2}e^{\nu/2} H_0 -\frac{p_r^\prime}{r}e^{(\nu-\lambda)/2} W -e^{\nu/2}S. \label{eq:LD_V}
\end{align}
\end{widetext}

Let us summarize how the above equations are derived. Equations~\eqref{eq:LD_H1} and \eqref{eq:LD_K} are the $(t,\theta)$-component and the $(t,r)$-component of Eq.~\eqref{eq:pEFE}. Equation~\eqref{eq:LD_W} can be obtained by combining the adiabatic relation Eq.~\eqref{eq:pGamma} with the $t$-component of Eq.~\eqref{eq:pSEC}.
It can also be derived with an alternative approach starting from the continuity equation as described in Appendix~\ref{app:continuity_eq}.
Equation~\eqref{eq:LD_X} is obtained by eliminating $H_0^\prime$, $K^\prime$ from the $r$-component of Eq.~\eqref{eq:pSEC} using the $(t,r)$ and $(\theta,r)$-components of Eq.~\eqref{eq:pEFE}.  The algebraic relation Eq.~\eqref{eq:LD_H0} is derived by eliminating the $H_0^\prime$, $K^\prime$ terms with $(t,r)$, $(r,r)$, and $(\theta,r)$-components of Eq.~\eqref{eq:pEFE}, and Eq.~\eqref{eq:LD_V} is found from the $\theta$-component of Eq.~\eqref{eq:pSEC}.  
Equation~\eqref{eq:dsigma} comes from the assumed relation
\begin{align}
    \delta \sigma = \frac{\partial \sigma}{\partial p_r} \delta p_r + \frac{\partial \sigma}{\partial \rho}\delta \rho + \frac{\partial \sigma}{\partial \mu} \delta \mu, \label{eq:dsigma}
\end{align}
which reduces to the Cowling limit if we take $\delta \mu = 0$.

To solve for the QNMs, we integrate Eqs.~\eqref{eq:LD_H1}--\eqref{eq:LD_X} numerically to obtain the perturbed solution of the NS interior. We begin the integration near the stellar center. Assuming the solutions are regular yields the following constraints:
\begin{widetext}
\begin{align}
    H_1^{(0)} =& \frac{16\pi}{\ell(\ell+1)} (\rho^{(0)}+p_r^{(0)}) W^{(0)} + \frac{2}{\ell+1}K^{(0)}, \label{eq:H1_expansion}\\
    X^{(0)} =& (\rho^{(0)}+p_r^{(0)}) e^{\frac{\nu^{(0)}}{2}}\Bigg\{\bigg[\frac{4\pi}{3}(\rho^{(0)}+3p_r^{(0)}) +\frac{2\sigma^{(2)}}{\rho^{(0)}+p_r^{(0)}}-\frac{\omega^2 e^{-\nu^{(0)}}}{\ell}\bigg]W^{(0)} + \frac{K^{(0)}}{2}\Bigg\}, \\
    S^{(0)} =& 0, \label{eq:dsigma_expansion}
\end{align}
\end{widetext}
where the superscript $(n)$ denotes the coefficient of the $n$th order expansion of the quantity in series of $r$ about the stellar center. At the surface, the vanishing Lagrangian pressure perturbation gives $X(R) = 0$. The interior solution is then matched with the metric perturbation in vacuum at the stellar surface. We follow the Zerilli's method described in \cite{Lindblom_1983} (also check \cite{Lu_2011} for amendments of several typos) to solve for the complex frequency corresponding to purely outgoing solutions at far-field region.

Note that Eq.~\eqref{eq:dsigma_expansion} combined with Eq.~\eqref{eq:dsigma} effectively places an extra constraint on $\sigma$ near the center: 
\begin{align}
    \left(\frac{\partial \sigma}{\partial x}\right)_\text{eq} =\mathcal{O}(r^2), \label{eq:dsigma_regular}
\end{align}
where $x$ represents ($p_r$, $\rho$, $\mu$). This is similar to how the background structure equation (Eq.~\eqref{eq:TOV_p}) restricts $\sigma$ to go as $r^2$ at the center (at least $r$ to avoid irregularity). The H anisotropy model is modified from the original form in~\cite{Horvat_2010} to satisfy the above condition.

\section{\label{sec:unstable_modes} Unstable quasinormal modes}

\begin{figure*}
\includegraphics[width = 8.6cm]{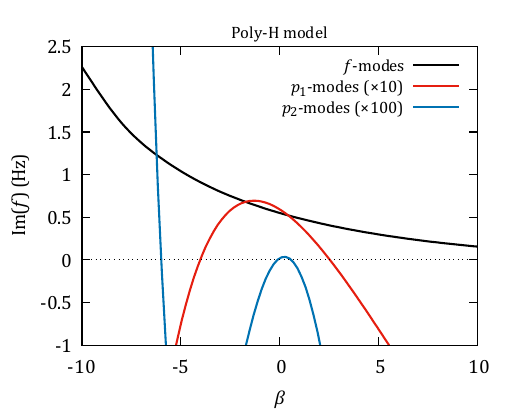}
\includegraphics[width = 8.6cm]{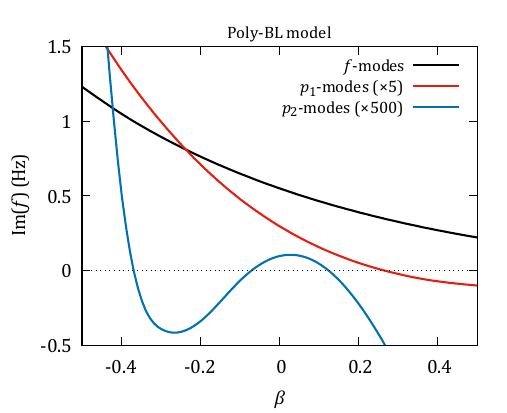}
\includegraphics[width = 8.6cm]{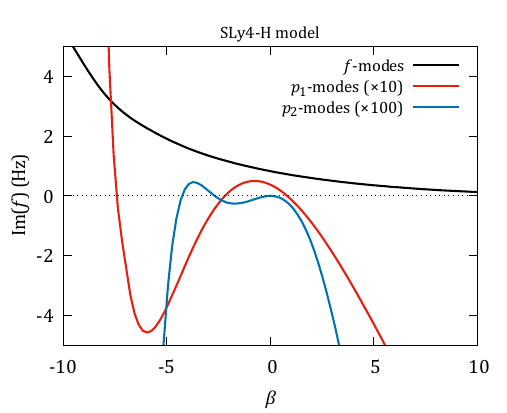}
\includegraphics[width = 8.6cm]{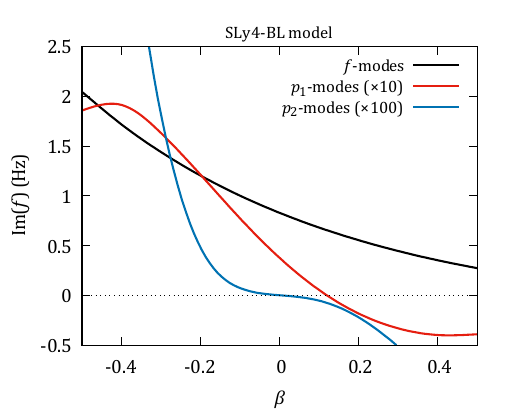}
\caption{The imaginary part of the first three $\ell=2$ QNM frequencies, $f = \omega/(2\pi)$, against the anisotropy parameter $\beta$ for different combinations of the EOSs (Poly vs SLy4) and the anisotropy models (H vs BL). The $p$-mode frequencies are scaled up by the constant factors specified in the legends to provide better visualization of the zero-crossings. The central densities of the Poly models and the SLy4 models are set at $7.455\times10^{14}\text{g cm}^{-3}$ and $9.88\times10^{14}\text{g cm}^{-3}$ respectively, such that the isotropic ($\beta = 0$) NS model has a mass of 1.40~$M_\odot$.}
\label{fig:imf_beta}
\end{figure*}

\begin{figure*}
\includegraphics[width = 8.6cm]{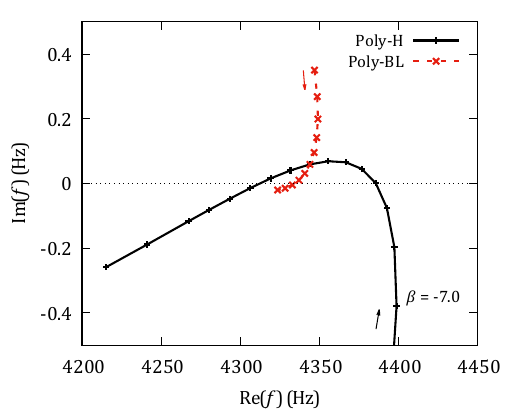}
\includegraphics[width = 8.6cm]{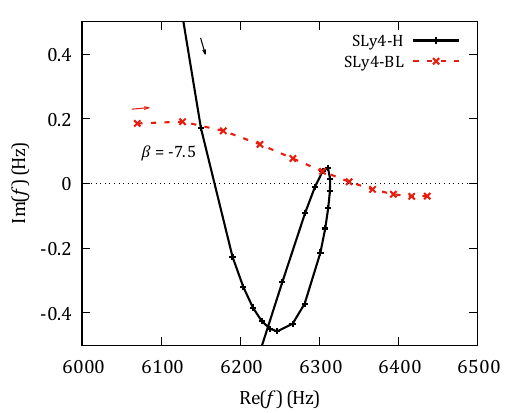}
\caption{The $\ell=2$ $p_1$-modes in the complex frequency plane of the models in Fig.~\ref{fig:imf_beta} for Poly (left) and SLy4 (right) EOSs. The arrow indicates the direction of increasing $\beta$, in which the absolute values are chosen to be $|\beta|\le 10$ and $|\beta|\le 0.5$ for the H-model and the BL-model respectively. The value of $\beta$ corresponding to the first point within the plotting range is shown for the SLy4-H model. Note that the curves intersect in the isotropic case (the intersection with Re($f$) = 6304~Hz for the right panel.}
\label{fig:complex_f}
\end{figure*}

Here we show the results of the $\ell=2$ QNMs of anisotropic NS models where the $p$-mode frequencies can have a negative imaginary part for a certain range of anisotropic parameters. These modes are interpreted as unstable QNMs. These results are obtained by numerically solving the linearized perturbed Einstein equations and stress energy conservation as described in Sec.~\ref{sec:formulation}.
The equations have been checked to correctly reduce to those in the Cowling limit of anisotropic stars given in \cite{Doneva_2012}. 
We performed our calculations using two independent numerical codes
and found consistent results with a difference of $\sim 1\%$. The major code is publically available at \cite{code}.

In Fig.~\ref{fig:imf_beta}, we present the imaginary parts of the QNM frequencies of the $f$-mode, $p_1$-mode, and the $p_2$-mode against $\beta$ for various combinations of the $\rho$--$p_r$ EOS (Poly and SLy4) and the anisotropy models (H and BL). 
First, the imaginary part of all modes in the isotropic limit ($\beta = 0$) is also positive, which serves as a sanity check of the numerical results, as the governing equations readily reduce to the isotropic case given in \cite{Detweiler_1985, Lu_2011}. 
Next, we see that the $f$-modes have positive imaginary frequencies for any anisotropy parameter $\beta$. 
On the other hand, the imaginary parts of the $p$-modes cross zero as we vary $\beta$ away from 0,  and become negative in a certain range of $\beta$. For the $p_2$-modes of the Poly-H model, it even crosses zero three times within the range of $\beta$ we consider.
We also found similar zero-crossing behaviors in $p$-modes at higher radial orders. This illustrates that there are multiple unstable $p$-modes within the anisotropic NSs. For a particular value of $\beta$, even the lowest order modes being stable does not necessarily imply a stable star. In principle, one needs to solve for an infinite number of $p$-modes to check for stability.

We further illustrate the instability in contrast to the isotropic case, in which a smooth sequence of QNMs is forbidden to cross the real axis \cite{Detweiler_1973, Ipser_1975}. We focus on the $p_1$-mode frequencies and present in Fig.~\ref{fig:complex_f} a set of anisotropic NS models with various $\beta$ in the complex frequency plane. We start increasing $\beta$ from the corresponding lower bound value in the direction indicated by the arrow in each sequence of models. The frequency crosses the real axis multiple times depending on the EOSs and anisotropy models. In the case of isotropic stars, these crossings are forbidden by the equation of motion \cite{Detweiler_1973} as we describe in more detail below.

We next perform a further analytic study by following \cite{Detweiler_1973} 
that was devised for variational principle. They showed, in the isotropic case, that the QNM frequency can be related to the eigenfunctions as an integral formula consisting of pairs of the perturbation variables, schematically written in the form
\begin{align}
    \omega^2 \int_0^\infty I_1 dr = \int_0^\infty I_2 dr + I_3(R) + I_4(\infty), \label{eq:integral_eq16}
\end{align}
where $\omega$ is the QNM frequency and $I_1$ to $I_4$ are sums of pairs of the perturbation variable eigenfunctions, which are in general complex (see Appendix~\ref{app:D73_integral} for the explicit expressions). The term $I_3$ is evaluated at the stellar radius $R$ while $I_4$ is evaluated at infinity. In $I_1$ to $I_3$, the perturbation variables are expressed in complex conjugate pairs such that they are guaranteed to be real. 
For $I_4$, the outgoing wave condition for QNMs makes its imaginary part non-vanishing, except when $\omega$ is purely imaginary. As a result, $\omega$ cannot be real unless there is no outgoing wave at infinity, meaning that the QNM frequencies cannot cross the real axis in the complex frequency plane except at the origin.

For the anisotropic case, we find the following extra terms on the right-hand side of Eq.~\eqref{eq:integral_eq16}: 
\begin{align}
-\omega^2 \int_0^R J_1 dr + \int_0^R J_2 dr + J_3(R), \label{eq:integral_anisotropic_terms}
\end{align}
where $J_1$ and $J_3$ are real, while $J_2$ is in general complex (see Appendix~\ref{app:D73_integral} for the explicit forms of these terms). Together with $I_4(\infty)$ in Eq.~\eqref{eq:integral_eq16}, there is the possibility of $\omega$ becoming purely real.
As a result, $\omega^2$ is allowed to cross the real axis for $\beta \neq 0$.

\section{\label{sec:conclusion} Discussions and conclusion}

In this work, we demonstrated that the $p$-modes can become unstable in some range of the anisotropy parameter $\beta$ with two different anisotropy models and two NS EOSs. We analytically illustrated that such instabilities are allowed to exist based on the mathematical structure of the eigenvalue problem, in contrast to the isotropic case. 
In Appendix~\ref{app:num_data}, we compared our calculations in full GR with those within the Cowling limit and found that the latter is valid for anisotropic NSs with an error of $< 10\%$ for $p$-modes. Additional data are also provided for numerical comparisons.
%

The implication of the instability still requires further analysis. We can expect the nonlinear effects will become important as the unstable mode amplitude grows exponentially with time, causing the linearized theories to eventually break down. 
A feedback mechanism is expected to explain the source of energy for the growth of QNMs while gravitational waves carry away the energy.
This should cause some permanent changes in the stellar background and/or the anisotropy model while extracting the energy for the growth. 

We can draw some insights from the Chandrasekhar-Friedman-Schutz (CFS) instability of a rotating star \cite{Chandrasekhar_1967, Friedman_1978}, in which a QNM is seen as a prograde mode in the inertial frame, but retrograde in the co-rotating frame, thus having opposite  (canonical\footnote{The importance of distinguishing the canonical and physical energy and angular momentum in $r$-modes is emphasized in \cite{Levin_2001})}) angular momenta (see e.g., \cite{Maggiore_2018}). When it loses positive angular momentum through radiation, it is driven to even larger amplitudes since the mode carries a negative angular momentum. In this case, the energy source that drives the instability comes from the rotation of the background star. Such an effect cannot be captured without considering the feedback on the background rotation. 

In the setup of this study, there is no background rotation to source the growth in the mode amplitude. However, some permanent changes to the background star are expected for the mode to increase its amplitude spontaneously. One possibility is that the anisotropy will decay in a short time due to certain feedback mechanisms from the unstable mode, which can only be obtained through calculations beyond linearized perturbation theories. This may restrict the form of the anisotropy models that can be sustained in a NS.

The effects of viscous damping within the NS are also neglected in this study. These effects are important in determining the suppression of the instability due to other dissipative processes. As a reference, the dominant (shear) viscosity damping rate of typical isotropic NSs is $\lesssim 10^{-8}$~Hz for $\ell = 2$ modes \cite{Cutler_1990}, which is much smaller than the growth rate of the unstable $p$-modes of the majority of the parameter range reported here. This means the known viscous damping effect does not suppress the instability.

This work opens up several new avenues in the study of anisotropy-driven instabilities.
First, to explore the physical origin of the instability, we need to identify the energy source for the growth of the mode amplitude. A variational approach may allow us to construct an energy balance law as in \cite{Detweiler_1973}.
Second, the stability of other QNMs, like the $f$-modes and $g$-modes, is also an interesting problem to consider. Although our numerical results suggest that the former are stable within the range of $\beta$ considered, it requires a thorough study with a larger range of NS parameters and anisotropy models to draw a meaningful conclusion. 
Finally, we can further explore the prospects of constraining the anisotropy models through the unstable QNMs by requiring stability.

Lastly, we re-emphasize that the form of the anisotropic pressure and its perturbation in this paper are parametrized models of assumed forms instead of being derived from fundamental thermodynamic principles, which can be unrealistic and are considered effective models only. The recent series of preprints \cite{Cadogan_2024_1, Cadogan_2024_2, Cadogan_2024_3} address the issues of these existing anisotropic models that enforce an explicit $r$-dependence to fix the problem of singularity at the stellar center. They point out that these anisotropic EOSs do not inform how matter perturbs and can violate the principle of equivalence. For this reason, they developed the theory of a relativistic anisotropic fluid starting from the action using a consistent model inspired by the theory of liquid crystals. This theory can be applied to non-radially symmetric configurations and allows a self-consistent derivation of the pulsation equations. Similar analyses of anisotropy from elasticity in relativity can be found in \cite{Karlovini_2003, Karlovini_2004}. As we discuss in Appendix~\ref{app:continuity_eq}, 
our formalism can be generalized by modifying the form of the perturbed anisotropy to describe anisotropy from different origins (e.g., elasticity, magnetic field, etc). It is worth exploring whether the instability reported in this paper exists for QNMs derived from such a theory, i.e., whether it emerges from some flawed assumptions in the anisotropy model or physically exists.

\emph{Note added.---} While we were working on this project, a preprint \cite{Mondal_2023} appeared on arXiv that provided a derivation of perturbation equations for oscillating anisotropic NSs in full GR. However, we noticed that the authors have incorrectly assumed an isotropic deformation of the fluid when applying the thermodynamic relations, while the deformation along each direction should be different for a generic nonradial deformation (see Eqs.~\eqref{eq:1st_law_aniso} and \eqref{eq:strain} of Appendix~\ref{app:continuity_eq}). Hence, the perturbation equations presented in~\cite{Mondal_2023} do not correctly reduce to those in the Cowling limit~\cite{Doneva_2012} and are inconsistent with those in this paper that were derived through a different approach than~\cite{Mondal_2023}. Furthermore, Ref.~\cite{Mondal_2023} used the original anisotropy model by Horvat \emph{et al}. \cite{Horvat_2010} that does not satisfy the correct boundary conditions for the perturbation equations. 
The main focus of this paper is on the stability of QNMs, which is also different from \cite{Mondal_2023}.


\vspace{2mm}
\begin{acknowledgments}
K.Y. acknowledges support from NSF Grant PHY-2339969 and the Owens Family Foundation. The authors thank Phil Arras for useful
discussions. The authors also thank the anonymous referee for providing important comments on the origin of the instability and pointing out interesting references about the CFS instability.
\end{acknowledgments}

\appendix

\section{An alternative derivation with the continuity equation} \label{app:continuity_eq}

In this section, we provide an alternative way to derive the pulsation equations, Eq.~\eqref{eq:LD_W} in specific. Instead of using the $t$-component of Eq.~\eqref{eq:pSEC}, we utilize the continuity equation of particle number density and the relation of thermodynamics. At the end, we explain how this method allows incorporating a more general form of the EOS in the perturbed configuration.

We begin by reviewing the derivation of the pulsation equations for \emph{isotropic} stars. It is common to introduce the particle number density, $n$, which is governed by the continuity equation (see, e.g., \cite{Friedman_1975}):
\begin{align}
    \nabla_\alpha (n u^\alpha) = 0
    \implies \frac{\Delta n}{n} = - \frac{1}{2}h^{\alpha \beta} \Delta g_{\alpha \beta} .\label{eq:continuity}
\end{align}
The number density is an independent variable in the EOS, i.e., $p =p(n)$, $\rho = \rho(n)$, but does not enter the equation of motion directly once we have $\rho$ as a function of $p$. The pressure $p$ here is the isotropic pressure. The Lagrangian perturbation of $n$ is related to that of $\rho$ through the thermodynamic relation
\begin{align}
    \Delta\rho = (\rho+p) \frac{\Delta n}{n}. \label{eq:1st_law_iso}
\end{align}
Combining Eqs.~\eqref{eq:continuity}, \eqref{eq:1st_law_iso} and the adiabatic relation $\gamma \Delta \rho/(\rho+p) = \Delta p/p$ gives the $t$-component of the stress energy conservation (Eq.~\eqref{eq:pSEC}) of an isotropic fluid.

While Eq.~\eqref{eq:continuity} still holds in the anisotropic case, Eq.~\eqref{eq:1st_law_iso} needs to be modified, since the work done on the fluid does not depend only on the volume change, but is directional dependent\footnote{Mondal and Bagchi \cite{Mondal_2023} incorrectly assumed that Eq.~\eqref{eq:1st_law_iso} is valid also for anisotropic stars.}. This requires us to explicitly write down the change in energy due to work done in each direction (see, e.g., \cite{Landau_1986, Karlovini_2004}, for the generalized thermodynamic relations in an anisotropic medium):
\begin{align}
    \Delta \rho = (\rho+p_r) \frac{\Delta n}{n} + \sigma (\Delta U^\theta{}_{\theta}+\Delta U^\phi{}_{\phi}), \label{eq:1st_law_aniso}
\end{align}
where $\Delta U_{\alpha \beta}$ is the perturbative Lagrangian strain tensor \cite{Carter_1972} \footnote{In fluid dynamics, it is more common to use the strain rate tensor to describe the deformation of fluid elements during the flow. Since each fluid element in the pulsating star oscillates about its equilibrium position, the strain rate just differs from the strain by a factor of $i\omega e^{-\nu/2}$.}, given by
\begin{align}
    \Delta U_{\alpha \beta} =\Delta h_{\alpha \beta} = \frac{1}{2} h_\alpha{}^\mu h_\beta{}^\nu \Delta g_{\mu \nu}, \label{eq:strain}
\end{align}
and $\Delta U^\alpha{}_{\beta} = g^{\alpha \mu} \Delta U_{\mu \beta}$.
Equation~\eqref{eq:1st_law_aniso}, substituted with the adiabatic relation, Eq.~\eqref{eq:pGamma}, reads
\begin{align}
    \frac{\Delta p_r}{p_r} = \gamma \left[\frac{\Delta n}{n} + \bar\sigma (\Delta U^\theta{}_{\theta}+\Delta U^\phi{}_{\phi})\right].
\end{align}
One can show that this is equivalent to Eq.~\eqref{eq:LD_W} using Eqs.~\eqref{eq:continuity} and \eqref{eq:strain}, which are explicitly written as
\begin{align}
    \frac{\Delta n}{n} =& \sum_{\ell, m}\Bigg[\frac{1}{2}H_0 + K - \frac{e^{-\lambda/2}}{r} W^\prime \nonumber\\
    &- \frac{\ell+1}{r^2} e^{-\lambda/2} W- \frac{\ell(\ell+1)}{r^2} V\Bigg] r^\ell Y_{\ell m} e^{i \omega t} ,\\
    \Delta U^\theta{}_{\theta}+\Delta U^\phi{}_{\phi} =& \sum_{\ell, m}\Bigg[-K + 2 e^{-\lambda/2} \frac{W}{r^2} \nonumber\\ &+\frac{\ell(\ell+1)}{r^2} V\Bigg] r^\ell Y_{\ell m} e^{i \omega t}.
\end{align}

Notice that, in general, the anisotropic fluid can have different proportionality constants in each direction depending on the anisotropy EOS. That is, 
\begin{align}
    \frac{\Delta p_j}{p_j} = -\sum_i \gamma_{ij} \Delta U^i{}_{i},
\end{align}
where $\gamma_{ij}$ denotes the proportionality constant in the stress-strain relation for the spatial indices $i, j$.
This suggests a further generalization of the pulsation equations described in the formulation section. An even more general form can include non-zero contributions from off-diagonal elements of $\Delta U^i{}_{j}$ like in the elastic materials. We leave such studies for future work.

\section{Integral relation of the eigenvalue and eigenfunctions}\label{app:D73_integral}

The eigenfrequencies and the eigenfunctions of the pulsation equations are related through the integral equations Eqs.~\eqref{eq:integral_eq16} and \eqref{eq:integral_anisotropic_terms}, where $J_1$ to $J_3$ in Eq.~\eqref{eq:integral_anisotropic_terms} vanish in the isotropic limit and are non-zero only inside the star.
In this section, we provide the explicit forms of the terms $I_1$ to $I_4$ and $J_1$ to $J_3$.

\begin{widetext}
The terms $I_1$ to $I_4$ are obtained from Eq.~(16) of \cite{Detweiler_1973}, and are explicitly written as:
\begin{align}
I_1 =& e^{\frac{\lambda-\nu}{2}} r^2 \left\{(\rho+p_r)\left[\frac{1}{r^4}\left|\tilde W\right|^2 + \frac{\ell(\ell+1)}{r^2}\left|\tilde V\right|^2\right] - \frac{\ell(\ell+1)}{16\pi e^{\lambda} r^2}\left|\tilde H_1\right|^2 - \frac{1}{16\pi}\left[\left|\tilde K\right|^2 + 2\operatorname{Re}\left(\tilde K \tilde H_0^*\right)\right]\right\},\\
I_2 =& e^{\frac{\lambda+\nu}{2}} r^2 \Bigg\{ -\frac{(\rho+p_r)\nu^\prime A}{2 e^\lambda r^4} \left|\tilde W\right|^2 + \frac{1}{\gamma p_r}\left|\tilde P\right|^2 + \left[\frac{\ell(\ell+1)}{16\pi r^2} -\frac{3(\rho+p_r)}{4}\right]\left| \tilde H_0\right|^2 - \frac{1}{16\pi e^\lambda}\left|\tilde K^\prime\right|^2 \nonumber\\
&+ \frac{\rho+p_r}{\gamma p_r}\operatorname{Re} \left(\tilde P \tilde H_0^*\right) - \frac{(\rho+p_r) A}{e^{\lambda/2} r^2} \operatorname{Re}(\tilde W \tilde H_0^*) + \frac{1}{8\pi e^\lambda}\operatorname{Re}\left[\left(\nu^\prime \tilde H_0 +\tilde H_0^\prime\right) \tilde K^{\prime \, *}\right]\Bigg\},\\
I_3 =& \frac{e^\nu \rho}{2} \left[2 e^{-\nu/2}\operatorname{Re}\left(\tilde W \tilde H_0^*\right) + \frac{\nu^\prime}{r^2}\left|\tilde W\right|^2\right],\\
I_4 =& \frac{e^\nu r^2}{16\pi} \left[\tilde K^*\left(\tilde H_0^\prime + \nu^\prime \tilde H_0 - \tilde K^\prime\right) - \tilde K^\prime \tilde H_0^*\right].
\end{align}
\end{widetext}
Here, $\tilde P$ is defined through $\delta p_r = \sum_{\ell, m} \tilde P Y_{\ell m} e^{i\omega t}$, and the symbol $\operatorname{Re}$ denotes taking the real part.
The perturbation variables are redefined in the following ways to simplify the expressions: 
\begin{align}
r^\ell K =& \tilde K, \quad r^\ell H_0 = \tilde H_0, \quad r^\ell V = -\tilde V, \nonumber\\ r^\ell S =& \tilde S,
r^{\ell+1} H_1 = \tilde H_1, \quad r^{\ell+1} W = \tilde W. \label{eq:tilde_var}
\end{align}
We see that $I_1$ to $I_3$ are real while $I_4$ is in general complex if we impose the purely outgoing boundary condition at infinity \cite{Price_1969, Detweiler_1973}.

The terms $J_1$ to $J_3$ are derived using the procedures described in Appendix A of \cite{Detweiler_1973}, with the Einstein equations and stress-energy conservation for the anisotropic case:
\begin{widetext}
\begin{align}
    J_1 =& - \ell(\ell+1) e^{\frac{\lambda-\nu}{2}}(\rho+p_r)\bar\sigma \left|\tilde{V}\right|^2, \label{eq:J1}\\
    J_2 =& e^{\frac{\nu}{2}}\Bigg[(\rho+p_r)\bar\sigma \left(-\tilde K^\prime + \frac{2e^{-\frac{\lambda}{2}}A}{r^3} \tilde W\right) - \frac{2 \bar\sigma}{r} \left(1+\frac{\rho+p_r}{\gamma p_r}\right)\tilde P  +\frac{2}{r}\tilde S\Bigg] \tilde W^* \nonumber \\
    &+e^{\frac{\nu+\lambda}{2}}\left(\bar\sigma \tilde P + \frac{\rho+p_r}{2}\bar\sigma \tilde{H}_0 - \tilde S\right)\left[r^2 \tilde K^*+\ell(\ell+1)\tilde V^*\right], \label{eq:J2}\\
     J_3 =& \frac{2e^{\nu}\rho }{r^3} \bar\sigma \left|\tilde W\right|^2. \label{eq:J3}
\end{align}
\end{widetext}


The perturbation variables of $J_1$ and $J_3$ are paired in the same way as $I_1$, while $J_2$ is in general complex as the complex perturbation variables cannot be paired in the same ways as $I_1$ to $I_3$. Although it is not obvious that the complex pulsation variables within $J_2$ cannot be paired up to form real terms, we verified this by numerically computing the integral using the solutions of the QNMs. 



Another perspective is to focus on $\tilde S$, which depends on the anisotropy ansatz itself. The choice of the functional form for $\sigma$ affects the perturbation variable $\tilde S$, which in general causes $J_2$ to be complex. 
In \cite{Hillebrandt_1976}, it is shown that the non-radial modes are stable as long as the anisotropy is finite in the Newtonian incompressible star with $\tilde S = \beta \tilde P$. We point out here that it holds specifically for this anisotropy ansatz and therefore does not contradict our findings in this paper. In addition, the instability comes from the imaginary part of the eigenfrequency, which is caused by the GR effect.
%

\section{Eigenfrequencies in Cowling limit}
\label{app:D73_integral_Cowling}

Here, we show that the instability reported in this paper does not occur under Cowling approximation by illustrating that the perturbation variables in $J_2$ (see Eq.~\eqref{eq:J2}) all pair up to be real when we drop the metric perturbations.


The full GR formalism can be reduced to the relativistic Cowling limit given in \cite{Doneva_2012}. Notice that the approximation procedure is not simply taking all metric perturbations to zero. The steps involved are the same as the isotropic case described in, e.g., \cite{Lindblom_1990, Zhao_2022}, where we set $H_0$, $K$, $H_1$ to zero in Eqs.~\eqref{eq:LD_W}, \eqref{eq:LD_X}, and \eqref{eq:LD_V}, and further drop the term $-4\pi (\rho+p_r)^2 e^{(\lambda+\nu)/2} W/r$ in Eq.~\eqref{eq:LD_X} \footnote{The reason for removing this term comes from Eq.~\eqref{eq:LD_K} in the Cowling limit, assuming $H_1$ does not reduce to zero in the same way as $H_0$ and $K$. An argument about $H_1$ being larger than the other metric perturbations in the Newtonian limit is given in \cite{Finn_1988}}. The resulting formulas are equivalent to Eqs.~(27), (29) and (30) of \cite{Doneva_2012} with a barotropic EOS, and can further be reduced to the isotropic limit \cite{McDermott_1983}.

Writing the Cowling approximation pulsation equations in terms of the variables with tilde introduced in Eq.~\eqref{eq:tilde_var}, we have
\begin{align}
    \tilde W^\prime =& \left(\frac{2 \bar\sigma}{r} - \frac{p_r^\prime}{\gamma p_r}\right) \tilde W - \frac{r e^{\lambda/2}}{\gamma p_r} \tilde P -\ell(\ell+1) e^{\lambda/2} (1-\bar\sigma) \tilde V, \label{eq:dW_tilde}\\
    \tilde P^\prime =& \frac{\rho+p_r}{r^2} e^{\lambda/2} \left[\omega^2 e^{-\nu} + \frac{\nu^\prime}{2} e^{-\lambda} A\right]\tilde W  \nonumber\\
    &- \left(1+\frac{\rho+p_r}{\gamma p_r}\right)\frac{\nu^\prime}{2} \tilde P - \frac{2}{r} \tilde S, \label{eq:dP_tilde}\\
    \tilde P =&  -(\rho+p_r)(1-\bar\sigma)\omega^2 e^{-\nu} \tilde V + \tilde S.  \label{eq:V_tilde}
\end{align}

Substituting Eqs.~\eqref{eq:dW_tilde}-\eqref{eq:V_tilde} into Eq.~\eqref{eq:J2} and setting all metric perturbations to zero, we can obtain
\begin{widetext}
    \begin{align}
        J_2 = -\left(e^{\frac{\nu}{2}}\tilde P \tilde W^*\right)^\prime + \omega^2 e^{\frac{\lambda-\nu}{2}} (\rho+p_r) \left[\frac{1}{r^2}\left|\tilde W\right|^2 - (1-\bar\sigma) \ell(\ell+1) \left|\tilde V\right|^2\right] + e^{\frac{\nu-\lambda}{2}} (\rho+p_r) \frac{\nu^\prime}{2} \frac{A}{r^2} \left|\tilde W\right|^2, 
    \end{align}
\end{widetext}
where the first term becomes a real boundary term after integration. Hence, combining $I_1$ to $I_3$ and $J_1$ to $J_3$ under Cowling approximation, $\omega^2$ must be real.

\section{Additional numerical data} \label{app:num_data}

\begin{figure*}
\includegraphics[width = 8.6cm]{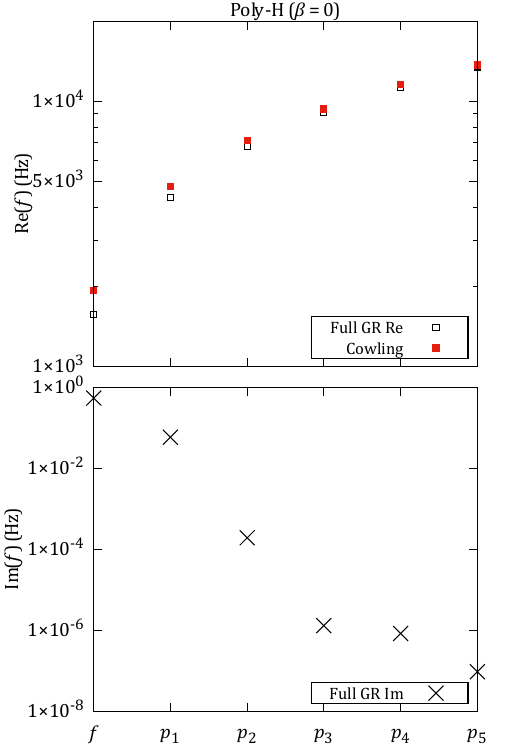}
\includegraphics[width = 8.6cm]{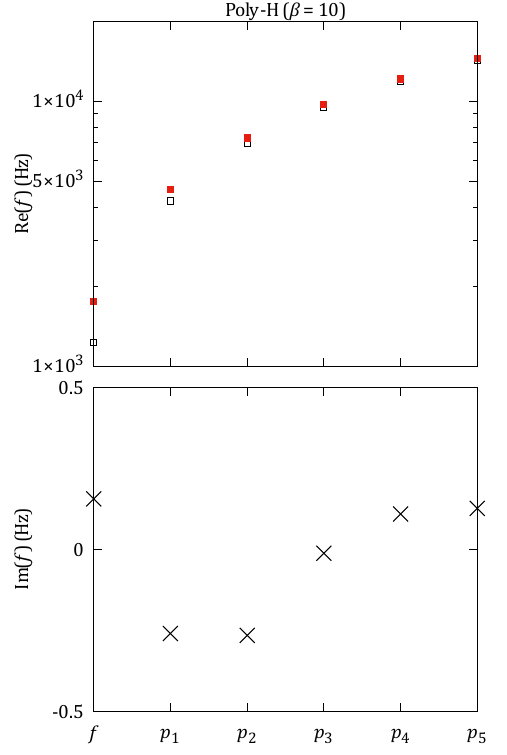}
\caption{The real (top panel) and imaginary (bottom panel) frequencies of the first six $\ell=2$ QNMs of the Poly-H model with central density 7.455$\times10^{14}~\text{g cm}^{-3}$. The left panel has $\beta = 0$ and the right panel has $\beta = 10$. We also present the normal mode frequencies computed using the relativistic Cowling approximation (Cowling) described in \cite{Doneva_2012}. 
}
\label{fig:f_rca}
\end{figure*}

In this section, we provide additional numerical data of the QNMs computed for the EOSs and anisotropy models shown in the main text. We demonstrate that the anisotropy has a small effect on the real part of the frequency, even though it changes the stability of the mode. We also quantify the validity of the Cowling approximation for computing the oscillation modes of anisotropic stars for the first time. 

In Fig.~\ref{fig:f_rca}, we show the real and imaginary parts of the QNM frequencies for the first six $\ell=2$ modes of the Poly-H model. 
The real part of the frequencies of the isotropic and the anisotropic case is of similar values for all the modes.
For the isotropic case, the imaginary part of the modes is all above zero and decreases with the mode order. Meanwhile, that of the anisotropic case does not show a monotonic behavior and can become negative.

\begin{table*}
\caption{\label{tab:mode_data} Numerical data of the first three $\ell=2$ QNM frequencies of the anisotropic NS models. The central densities of the Poly models and the SLy4 models are set at $7.455\times10^{14}\text{g cm}^{-3}$ and $9.88\times10^{14}\text{g cm}^{-3}$ respectively.}
\begin{ruledtabular}
\begin{tabular}{ccccccccc}
Model   & $\beta$ & $M$ ($M_\odot$) & \multicolumn{3}{c}{Re($\omega$) (kHz)} & \multicolumn{3}{c}{Im($\omega$) (Hz)}               \\
        &         &                 & $f$         & $p_1$       & $p_2$      & $f$   & $p_1$                & $p_2$                \\ \hline
Poly    & 0       & 1.40            & 1.574       & 4.344       & 6.784      & 0.550 & $5.93\times10^{-2}$  & $1.95\times10^{-4}$  \\
Poly-H  & $-10.0$   & 2.18            & 1.945       & 4.364       & 6.104      & 2.262 & $-1.573$               & 1.212                \\
Poly-H  & 10.0    & 1.10            & 1.233       & 4.215       & 6.948      & 0.156 & $-0.259$               & $-0.265$               \\
Poly-BL &$ -0.5$    & 1.74            & 1.791       & 4.347       & 6.561      & 1.231 & 0.351                & $7.80\times10^{-3}$  \\
Poly-BL & 0.5     & 1.16            & 1.325       & 4.324       & 6.931      & 0.221 & $-2.01\times10^{-2}$ & $-3.59\times10^{-3}$ \\
SLy4    & 0       & 1.40            & 1.934       & 6.304       & 9.348      & 0.830 & $3.77\times10^{-2}$  & $2.25\times10^{-5}$  \\
SLy4-H  & $-10.0$   & 2.52            & 2.527       & 5.715       & 7.672      & 5.561 & 13.24                & $-11.78$               \\
SLy4-H  & 10.0    & 1.06            & 1.322       & 6.147       & 8.877      & 0.129 & $-1.082$               & $-0.361$               \\
SLy4-BL & $-0.5$    & 1.75            & 2.226       & 6.070       & 9.029      & 2.046 & 0.186                & 0.121                \\
SLy4-BL & 0.5     & 1.16            & 1.556       & 6.437       & 9.109      & 0.274 & $-3.90\times10^{-2}$ & $-1.24\times10^{-2}$ \\
\end{tabular}
\end{ruledtabular}
\end{table*}

In Fig.~\ref{fig:f_rca}, we also show the frequencies obtained by the Cowling approximation. The $f$-mode frequencies show 20-30\% deviations from the full GR real frequencies. The percentage deviation decreases if we go to higher radial order $p$-modes. In the case of the $p_5$-modes, the deviation reduces to a $1\%$ level.


More numerical data are provided in Table.~\ref{tab:mode_data}, including the real and imaginary frequencies of the first three $\ell=2$ QNMs, as a reference for readers interested in a numerical comparison. The $\beta = 0$ case of the SLy4 EOS has been checked against numerical data available (e.g., \cite{Kunjipurayil_2022}), and attains good agreement. We see that with the same $\rho$--$p_r$ relation, the real part of the frequencies is not very sensitive to the change in $\beta$, while the imaginary part can change by orders of magnitude and even change the sign.


\bibliography{main}

\end{document}